# Monitoring of the terrestrial atmospheric characteristics with using of stellar and solar photometry

*Alekseeva G. A. (1), Galkin V. D. (1), Leiterer U (2), Naebert T. (2), Novikov V. V. (1), and Pakhomov V. P. (1).*

1) **Pulkovo Observatory (Saint-Petersburg, Russia).**
   **E-mail: alekseeva@gao.spb.ru**
2) **Meteorological Observatory Lindenberg (Germany).**
   **E-mail: ulrich.leiterer@dwd.de**

## Summary

On the basis of experience acquired at creation of the Pulkovo Spectrophotometric Catalog the method of investigation of a terrestrial atmospheric components (aerosols and water vapor) in night time are designed. For these purposes the small-sized photometers were created. Carried out in 1995-1999г.г. series of night and daily monitoring of the atmospheric condition in Pulkovo, in MGO by A.I.Voejkov., in Germany (complex experiments LITFASS 98 and LACE 98) confirmed suitability of devices, techniques of observations and their reduction designed in Pulkovo Observatory for the solution of geophysical and ecological problems. A final aim of this work - creation of small-sized automatic complexes (telescope + photometer), which would be rightful component of meteorological observatories. Such complexes will work without the help of the observer and would provide the daily monitoring of a terrestrial atmosphere.

## Introduction

During last 25 years in Pulkovo Observatory the work for creation of stellar spectral energy distributions catalog in range 0.3-1.1 μm, [1], had been carried out for both hemispheres. Observations were received in different expeditions of Academy of Sciences (Armenia, Pamirs, Chile, Bolivia). The methods of fundamental absolute spectrophotometry were used for this purpose.

One of most important step was the careful investigation of atmospheric extinction in the moment of observation for whole spectral range mentioned above (including telluric bands $O_3$, $O_2$, $H_2O$). Some special methods was developed by us for this aim. As the result we received the large volume of data about behavior of night atmosphere in different places of Earth.

These data and methods aroused geophysicists' interest in its application to solution of geophysical tasks, because there were no any indirect methods for investigation of night atmosphere.

Thus our collaboration with some geophysical institutes of Russia (Arctic and Antarctic Research Institute) and Germany (Meteorological Observatory Lindenberg (MOL) and Institute of Polar and Mariner Researches by A. Wegener) was started. During last 7 years some remarkable results was received on this way [2,3,4,5].

## 1 Used Instruments and Method

For star-observations in Pulkovo Observatory two variants of starphotometer (PSP-94 and PSP-94SPCM) was used. These starphotometers was developed and created in Pulkovo Observatory during 1993-95 on the base of two receivers – photomultiplier FEU-112 for PSP-94 and counter SPCM-LQ100 (avalanche photodiode) for PSP-94SPCM. Both variants of photometer have the set of 10 quasi-monochromatic interference filters for range 0.39-1.05 **μm**.

For sun-observations in MOL the automatic sunphotometer ROBAS30 was used. For star-observations we used the new semi-automatic version of starphotometer made by MOL Dr. Shulz & Partner GmbH and Pulkovo Observatory jointly in 1997 (on the base of counter SPCM-LQ100). Both photometers have the set of quasi-monochromatic interference filters. For determinations of aerosol optical depths we used the following (**wavelengths in μm**):
**0.3991, 0.4512, 0.4922, 0.5510, 0.6538, 0.7915, 0.8553, 1.0539 (sunphotometer);**
**0.3907, 0.4444, 0.5007, 0.5328, 0.6019, 0.6729, 0.7786, 0.8629, 1.0452 (starphotometer).**
For determinations of water vapor column content filters **0.9517** and **0.9446** was used (for sun- and starphotometry respectively). For all filters the detailed curves of transmission was received in optic laboratory MOL and LOMO (for Pulkovo photometers).The time-interval between observations is 10 min for sunphotometry and about 3-5 min for starphotometry.

Here we present only main formulae and expressions with short method description.

$$m_\lambda = -2{,}5\lg U_\lambda \qquad \textbf{(1)}$$

where $U_\lambda$ - signal from object measured (voltages or imp/sec) in wavelength $\lambda$

$$m_\lambda^{obs.} = m_\lambda^{0} + \alpha_\lambda \cdot F(z) \qquad \textbf{(2)} \quad \textbf{(Bouger-Lambert' law),}$$

$m_\lambda^{obs.}$ - observable star magnitude

$m_\lambda^{0}$ - extraterrestrial instrumental star magnitude

$\alpha_\lambda$ - total extinction for air mass **one**

**F(z)** - air mass

Extinctional coefficient $\alpha_\lambda$ and geophysical total optical thickness $\delta_\lambda$ are connected by simple expression:

$$\alpha_\lambda = -2{,}5/\ln 10 \cdot \delta_\lambda = 1.085736 \cdot \delta_\lambda \qquad \textbf{(3)}$$

Coefficient of extinction $\alpha_\lambda$ in the expression **(2)** includes continuous components of extinction and equals:

$$\alpha_\lambda = \alpha_\lambda^{Ray.} + \alpha_\lambda^{aer.} + \alpha_\lambda^{ozon.} + \alpha_\lambda^{NO_2.} \qquad \textbf{(4)}$$

where $\alpha_\lambda^{Ray.}$, $\alpha_\lambda^{aer.}$, $\alpha_\lambda^{ozon.}$, $\alpha_\lambda^{NO_2}$ - Rayleigh, aerosol, ozone, and nitrogen dioxide components of extinction accordingly. If we calculated $\alpha_\lambda^{Ray.}$ and received $\alpha_\lambda^{ozon.}$, $\alpha_\lambda^{NO_2}$ from independent sources, it's possible to determine $\alpha_\lambda^{aer.}$ ($\delta_\lambda^{aer.} = \alpha_\lambda^{aer.}/1.085736 \cdot \delta_\lambda^{aer.}$; $\delta_\lambda^{Ray.} = P/1013.25 \cdot 0.00874 \cdot \lambda^{-(3.916+0.072\lambda+0.05/\lambda)}$, **P** - pressure in **hPa**):

$$\alpha_\lambda^{aer.} = \alpha_\lambda - (\alpha_\lambda^{Ray.} + + \alpha_\lambda^{ozon.} + \alpha_\lambda^{NO_2}) \qquad \textbf{(5)}$$

For water vapor band the expression **(2)** are transformed to

$$m_\lambda^{obs.} = m_\lambda^{0} + \alpha_\lambda \cdot F(z) + \Delta m(W) \qquad \textbf{(6)}$$

where

**Δm (W)** - water vapor absorption in star magnitudes

**W** - water vapor content on the line of view (in **cmppw**)

For absorption magnitude we used empirical model developed in [**6,7**]:

$$\Delta m(W) = c_\lambda \cdot [W_0 \cdot F(z)]^{\mu_\lambda} \qquad \textbf{(7)}$$

where $C_\lambda$ and $\mu_\lambda$ are empirical parameters, $W_0$ - water vapor content in atmosphere. Spectral parameters $C_\lambda$ and $\mu_\lambda$ received from laboratory investigations in Pulkovo Observatory [**8**] may be used for calculation of integral parameters for any filter with known spectral transmission curve.

In order to use expression **(2)** for determination $\alpha_\lambda$ directly we must know the individual spectral instrumental extraterrestrial magnitudes $m_\lambda^{0}$ for selected object.

If the source of radiation is only one (Sun), we have only one way to receive such magnitudes, namely to execute the set of observations in region with vary stable extinction ($\alpha_\lambda = const.$). For sunphotometer such observations was executed at Zugspitze (2970 m above sea level, German Alps), and solar individual spectral instrumental extraterrestrial magnitudes were received for all filters. One must take into account that magnitude for „water“ filter will be wrong (**Bouger-Lambert' law** use is wrong in this case).

If we have many sources of radiation (stellar observations) we may use so-called **Δ-method** (or Two Star Differential Method) as **the first approximation** for $\alpha_\lambda$ -determinations. For two stars with **$\Delta F(z)\geq 1$** we have from expressions **(2)**:

$$\alpha_\lambda = \frac{\left({}^2m_\lambda^{obs.} - {}^1m_\lambda^{obs.}\right) - \left({}^2m_\lambda^{0} - {}^1m_\lambda^{0}\right)}{F_2(z) - F_1(z)} \qquad \textbf{(8)}$$

So in this case we do not need to know individual extraterrestrial instrumental magnitudes. It is sufficient to know only the difference of extraterrestrial magnitudes for these stars. Such difference may be received from any homogeneous spectrophotometric catalog because for monochromatic (or quasi-monochromatic) radiation this difference is constant for whichever system. We used the Pulkovo Spectrophotometric Catalog [1] as most homogeneous.

When the first values $\alpha_\lambda$ (for mean moments between two stars observations) are received according expression **(8)**, we can receive for all filters the time-dependencies $\alpha_\lambda(\mathrm{UTC})$ during one night by means of polynomial approximation. Then the individual $\alpha_\lambda$ -values may be calculated from this time-dependencies, and the individual instrumental extraterrestrial magnitudes $m_\lambda^0$ may be received for every stellar observation. And finally mean values $\overline{m_\lambda^0}$ may be received by meaning for individual $m_\lambda^0$ for every star used. These $\overline{m_\lambda^0}$ may be used directly for $\alpha_\lambda$ -determinations from expression **(2).** The procedure described above we call **„the second approximation“**.

For water vapor content determinations the procedure is more complicated. In order to use expressions **(6)** and **(7)** for **$W_0$**-determinations we must know not only $m_\lambda^0$, but parameters **$C_\lambda$** and **$\mu_\lambda$** too. As the first approximation we may use the parameters calculated from spectral laboratory data **[8]** both for sun- and starphotometers' „water“ filters.

For solar observations we are forced to use as the first approximation the wrong value $m_\lambda^0$ received according to **Bouger-Lambert' law**. Then we received the first values **$W_0$** from expressions **(6)** and **(7)**.

For stellar observations we can receive the first values **$W_0$** from **Δ-method**:

$$W_0 = \left[\frac{\left({}^2m_\lambda^{obs.} - {}^1m_\lambda^{obs.}\right) - \left({}^2m_\lambda^{0} - {}^1m_\lambda^{0}\right) - \alpha_\lambda \cdot \left(F_2(z) - F_1(z)\right)}{C_\lambda \cdot \left(F_2(z)^\mu - F_1(z)^\mu\right)}\right]^{\frac{1}{\mu}} \qquad \textbf{(9)}$$

$\alpha_\lambda$ for „water“ filters are calculated by logarithmic interpolation between nearest filters outside water vapor band or from power approximation of data from all other filters.

Then we may make „the second approximation“ for **$W_0$** and receive values $\overline{m_\lambda^0}$ in „water“ filter for all stars used from expressions **(6)** and **(7)**.

With received values $\overline{m_\lambda^0}$ and with „laboratory“ parameters **$C_\lambda$** and **$\mu_\lambda$** all data of stellar observations was recalculated. Then we executed the calibration procedure with using radiosondes' $W_0$ -data in order to define „real“ parameters **$C_\lambda$** and **$\mu_\lambda$**, which give the best agreement. The calibration curve for all observations with starphotometer is shown on **Fig. 1**. It is power approximation of the dependence **$\Delta m\ (W)$** where **$W = W_{RS80}\ F(z)$** - water vapor content on the line of view (in **cmppw**). As result we received parameters:

$$C_\lambda = 0.^m582 \quad \text{and} \quad \mu_\lambda = 0.548 \qquad \textbf{(10)}$$

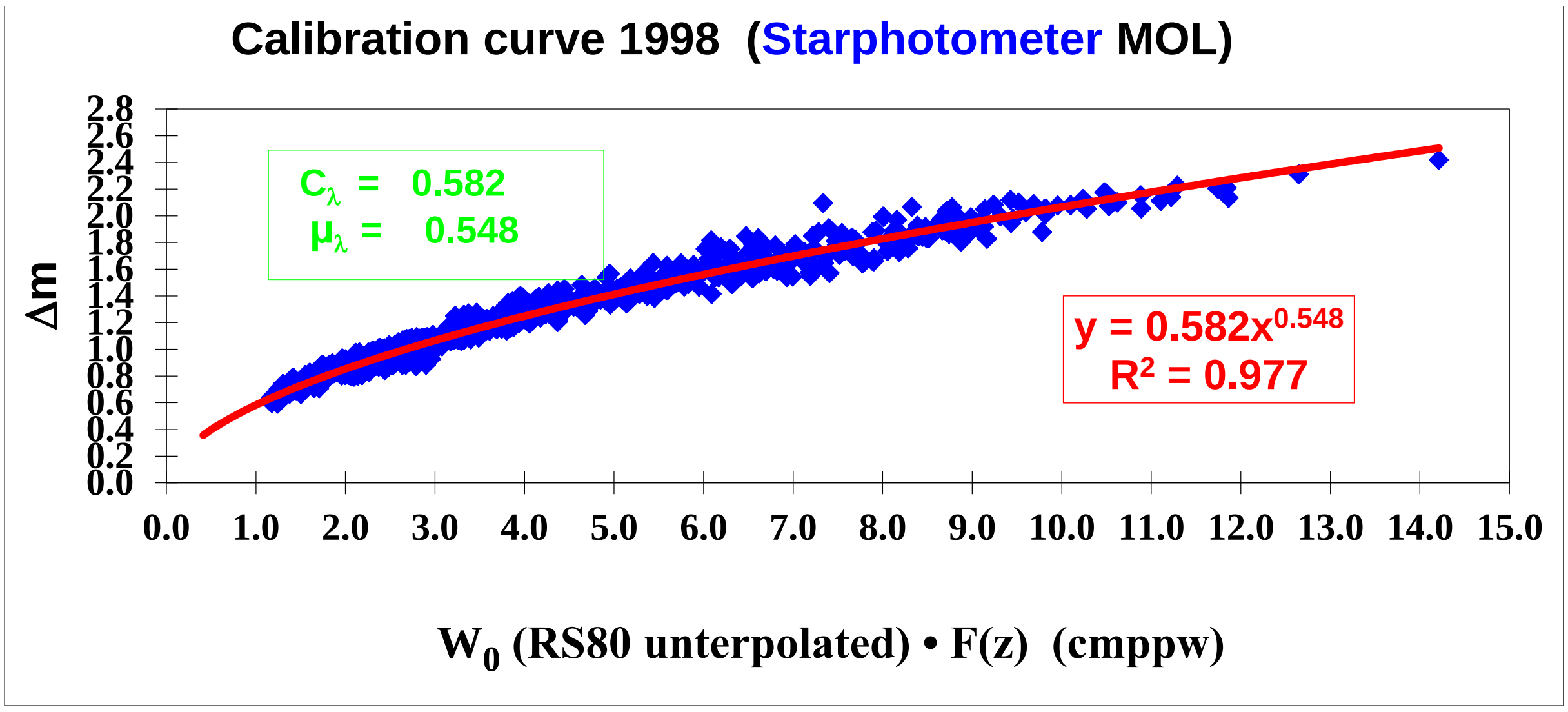

**Fig.1 Calibration curve for starphotometer MOL (1998).**

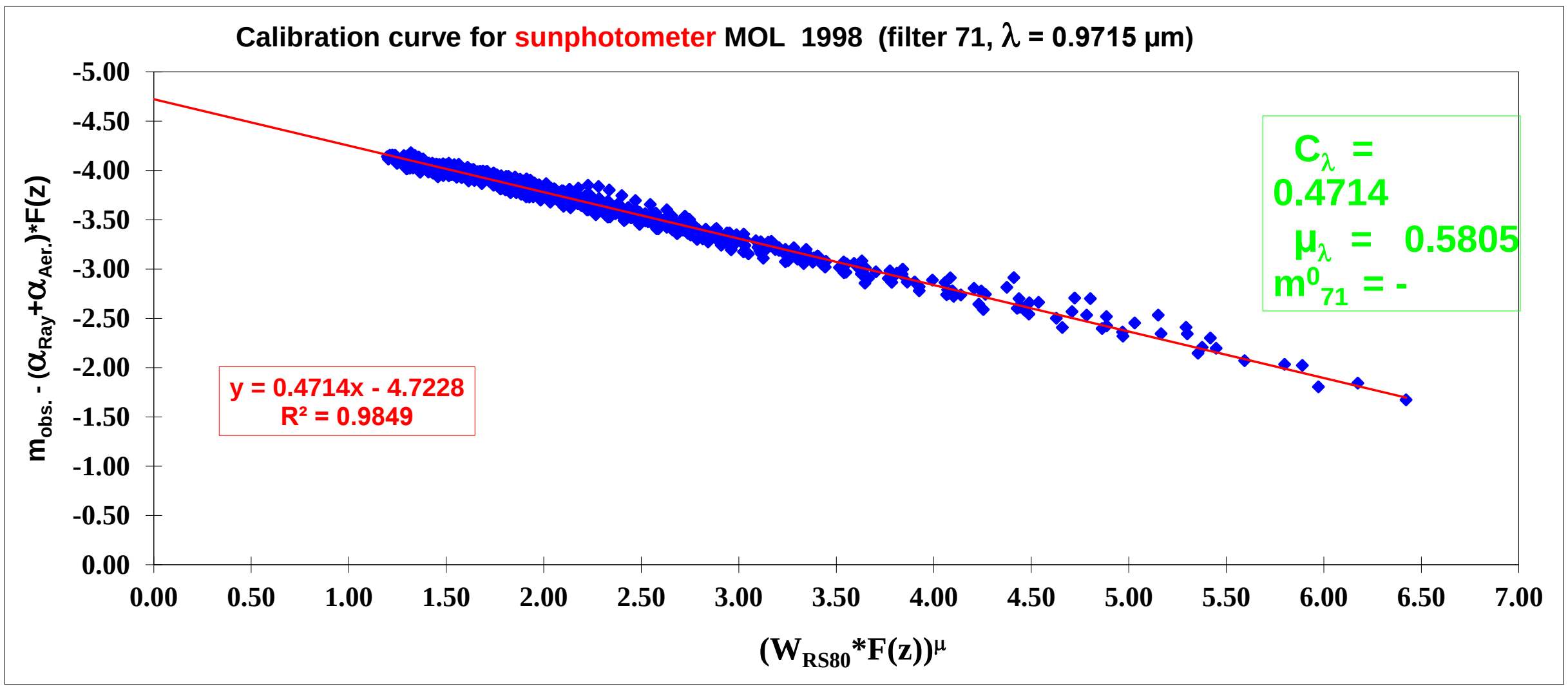

**Fig. 2. Calibration curve for sunphotometer MOL (1998).**

For sunphotometer we used the another dependence:

$$m_\lambda^{obs.} - (\alpha_\lambda^{Ray.} + \alpha_\lambda^{aer.}) \cdot F(z) = c_\lambda \cdot [W_{0_{RS80}} \cdot F(z)]^{\mu_\lambda}$$

The left part of dependence are known from observations, parameter $\mu_\lambda$ was determined by us from „laboratory“ data. Hence, by constructing of probabliest straight line, we can receive on crossing with axis of ordinates extraterrestrial instrumental solar magnitude $m_\lambda^0$, and the inclination of straight line will give the meaning of parameter $\mathbf{C_\lambda}$ specified on **RS80**-data. This calibration curve for all observations with **sunphotometer** is shown on **Fig. 2**. As result we received the following values:

$$\mathbf{C_\lambda = 0.^m4714\,,\quad \mu_\lambda = 0{,}5805 \quad \text{and} \quad m_\lambda^o = -\,4.7228 \qquad (11)}$$

All observational data for both photometers were recalculated with parameters **(10)** and **(11)**, and the final meanings of $\mathbf{W_0}$ were received.

## 2 Session of a daily monitoring of aerosol components of an atmosphere in Pulkovo Observatory and Main Geophysical Observatory by A. I. Voejkov.

On the basis of a method mentioned above the cycle of night photometric observations (August - September 1997) is spent with photometer PSP-94 in Pulkovo in parallel with solar photometric observations which carried out on traditional geophysical method by the employees of Arctic and Antarctic RI on territory of Main Geophysical Observatory by A. I. Voejkov. The observations were carried out in 10 selected narrow photometric bands in range of spectrum 0.380-1.050 μm. For the first time it has allowed to receive for elected dates samples of daily monitoring of condition of Earthly atmosphere over St.-Petersburg (dust, aerosols, and water vapor for the night hours). The day time and night data have shown good convergence. Examples received of temporary and spectral change the aerosol components of atmosphere are represented on **Fig. 3**.

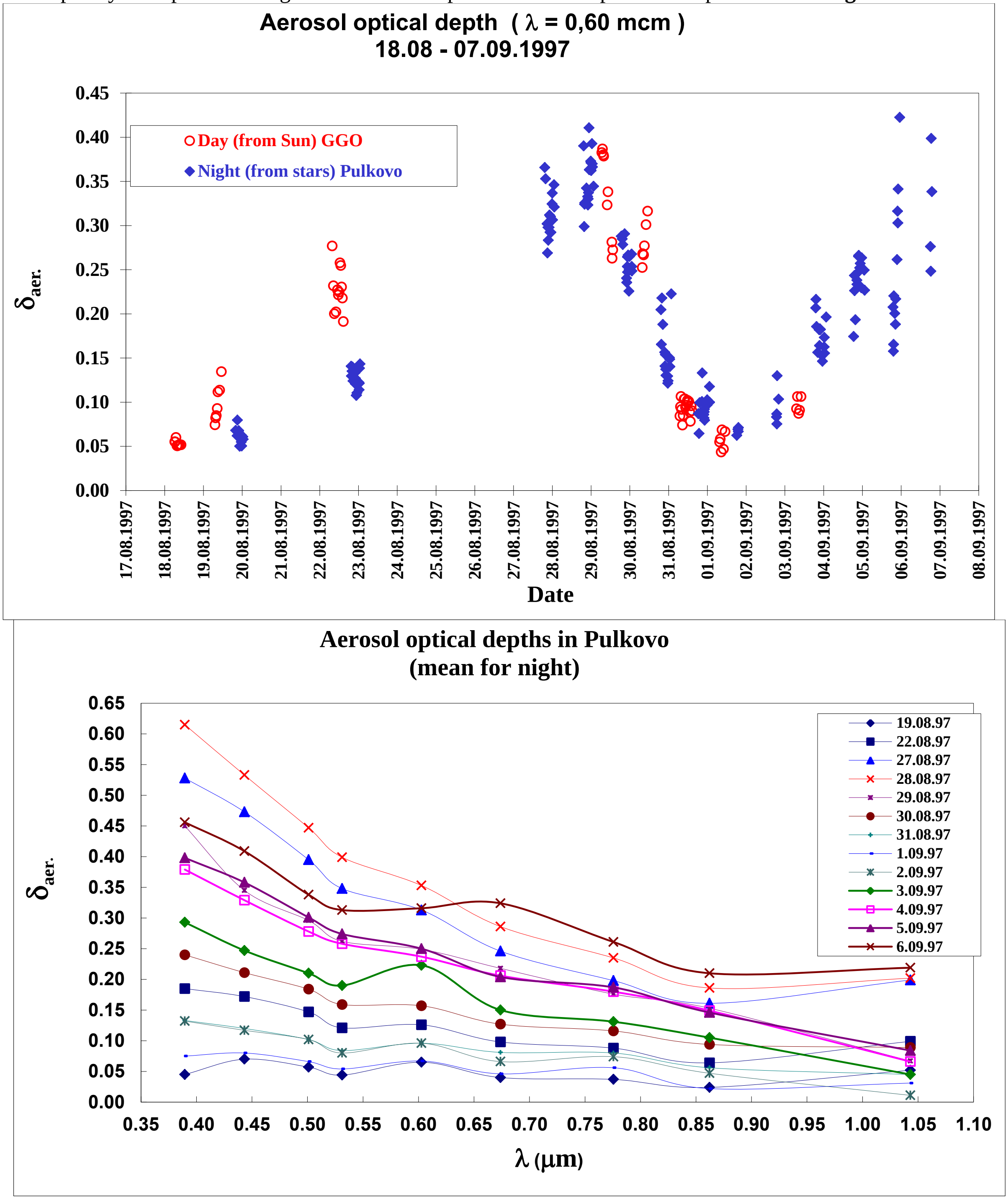

**Figure 3. Daily monitoring for Pulkovo and GGO (1997).**

## 3 Monitoring of aerosol components on station Koldewey (Spitsbergen).

Since 1994 on European station Koldewey (Spitsbergen) the regular (in spring-autumn period and all polar night) aerosol monitoring is carried out by employees of Institute of Polar and Marine Researches by A. Wegener (Germany) on method offered by Pulkovo Observatory with using of photometer, similar to MOL's photometer. In spring-autumn period solar observations are conducted in parallel. The example of such daily monitoring is represented on **Fig. 4** from [5].

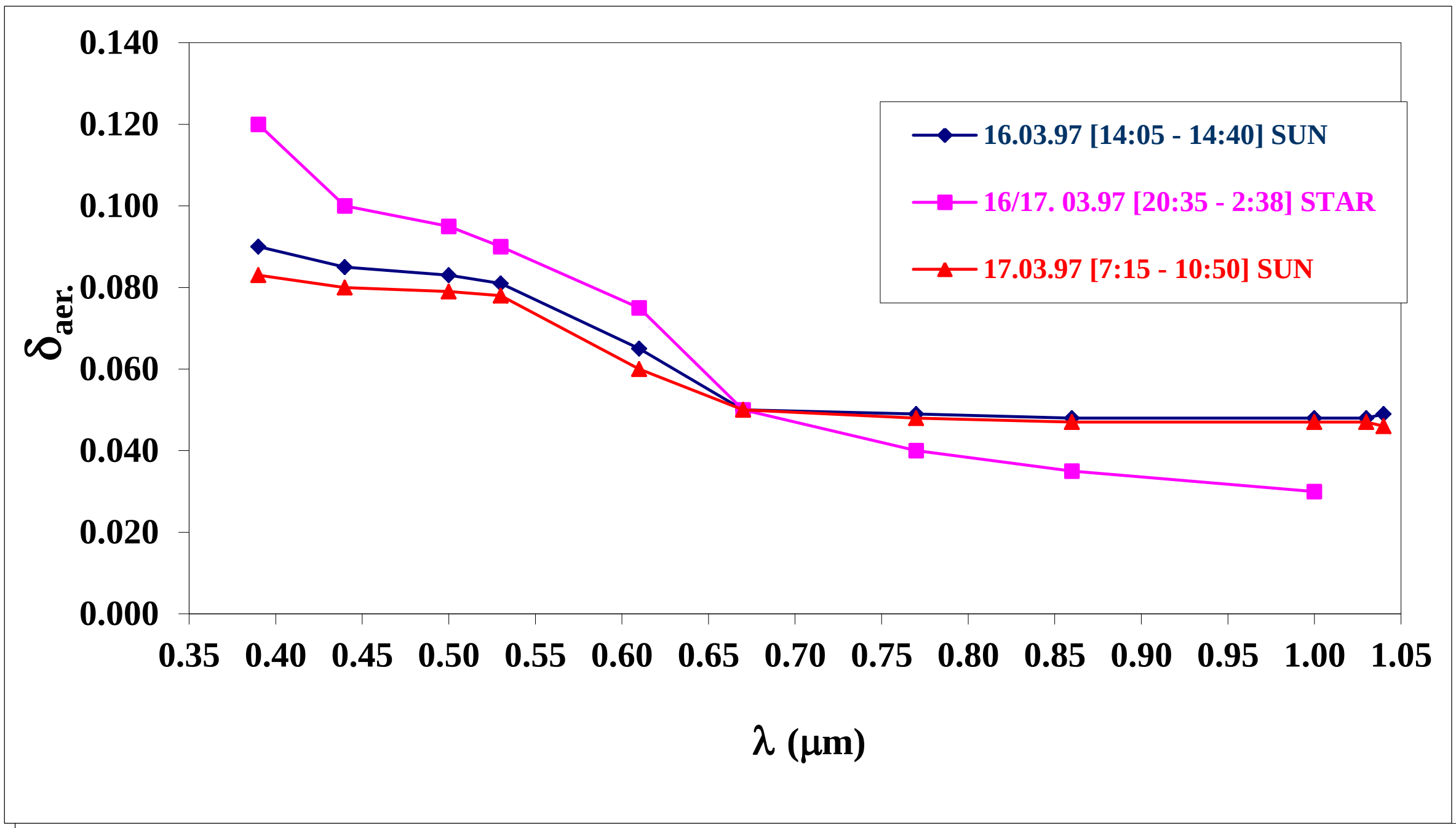

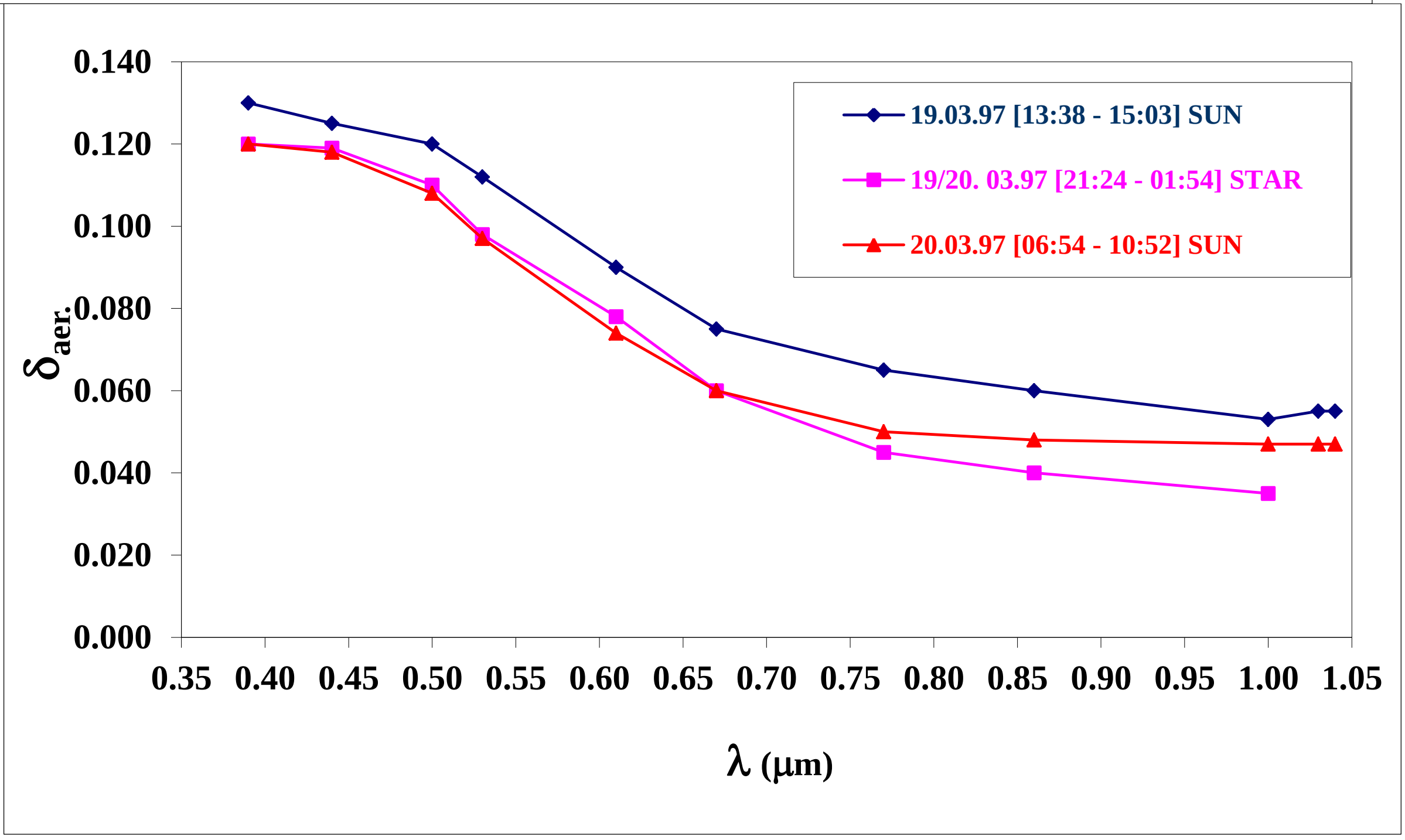

**Fig. 4.Sun- and Star photometer measurements on Koldewey station (March 1997)**

## 4 LITFASS 98 and LACE 98 experiments.

During summer 1998 two large-scale complex experiments LITFASS 98 (25.05 - 22.06) and Lindenberger Aerosol-Charakteristierungs-Experiment - LACE 98 (13.07 - 14.08) was executed in Meteorologisches Observatorium Lindenberg (MOL). The aim of both experiments was the complex daily monitoring of condition of Earthly atmosphere with determination of its fundamental meteorological properties and parameters in all vertical column over Lindenberg (Lindenberg's Column). About 20 research institutes of Germany and Pulkovo Observatory (Russia) had accepted participation in experiments, and the whole exiting set of geophysical devices and methods was used practically, including LIDAR's, microwave radiometer, radiosondes and whole complex of MOL-devices in Lindenberg and Falkenberg, meteo-aerostat, measurements from the plane and so on. For the first time star- and sunphotometers of MOL were used as the equal in rights devices alongside with other geophysical tools. The observations with both photometers were carried out practically each day and night, except absolutely overcast. Observational data were processed at once by cycle of programs developed in Pulkovo Observatory, and the results of processing (daily variations of aerosol optical depths and water vapor column content) were representing on daily briefings. Here we represent the summary of these results and it's comparison with radiosonde and microwave radiometer data.

As the result we have $\delta_\lambda^{aer.}$ - and **PW**-monitoring data for periods 26.05-22.06.98 and 14.07-16.08.98. The set of $\delta_\lambda^{aer.}$-data consists of individual spectral values for all „aerosol channels" and interpolated data for $\lambda = 0.550$ and 1.00 μm received by power approximation ($\delta_\lambda^{aer.} = A \cdot \lambda^{-B}$) for sunphotometry and by logarithmic interpolation for starphotometry.

For „water channels" we have data received by three different methods:

- **radiosonde** (**RS 80**, A-Humicap) relative humidity profiles, every **6 hours**;
- **optical methods** (according to expressions **1-11**) for sun- and starphotometry (every **10 min** for sunphotometer and every **3-5 min** for starphotometer);
- microwave radiometer's data (**microwave radiometer WVR-1100** [9]; $\nu$=23.8 GHz on; 31.4 GHz off every **minute**; running mean values of **10 min**).

.

The mean accuracy of sun- and starphotometry both for $\delta_\lambda^{aer.}$ - and **PW**-data is about **2-3%** (for conditions with vary unstable extinction it may decrease to **5-7%**). The detail analysis of data, accuracy and ways of accuracy increasing will be discussed in our special papers (in **Contrib. Atmos. Phys.**).

All received data are accessible as **Excel 5.0**-files in **MOL** (Dr. **U. Leiterer**, E-mail: **ulrich.leiterer@dwd.de).**

The fragment (for 4-11 Aug 1998) of resulting time series for $\delta_\lambda^{aer.}$ - and **PW**- data is shown on **Fig. 5**. This period was characterized by good atmospheric stability and respectively lower aerosol concentrations especially for 3-12 August ("golden" period). The "absolute" minimum of aerosol concentration was observed at night 10/11 August. For this night $\delta_\lambda^{aer.}$ was sometimes no more 0.01, or aerosol was absent in range of accuracy for some wavelengths. It was caused by a thin dry aerosol layer near 3 km height transported from forest fires in North America (British Columbia) to Central Europe.

One can see that as the rule we have very good agreement (difference is no more **5%**) for **PW**-data received by three methods. Only sometimes difference increases to **10%**.

The important feature is that the individual **PW**- time-variations are the same topologically for microwave radiometer's, sun- and starphotometric data. It means that we can investigate systematically the real rapid **PW**- time-variations.

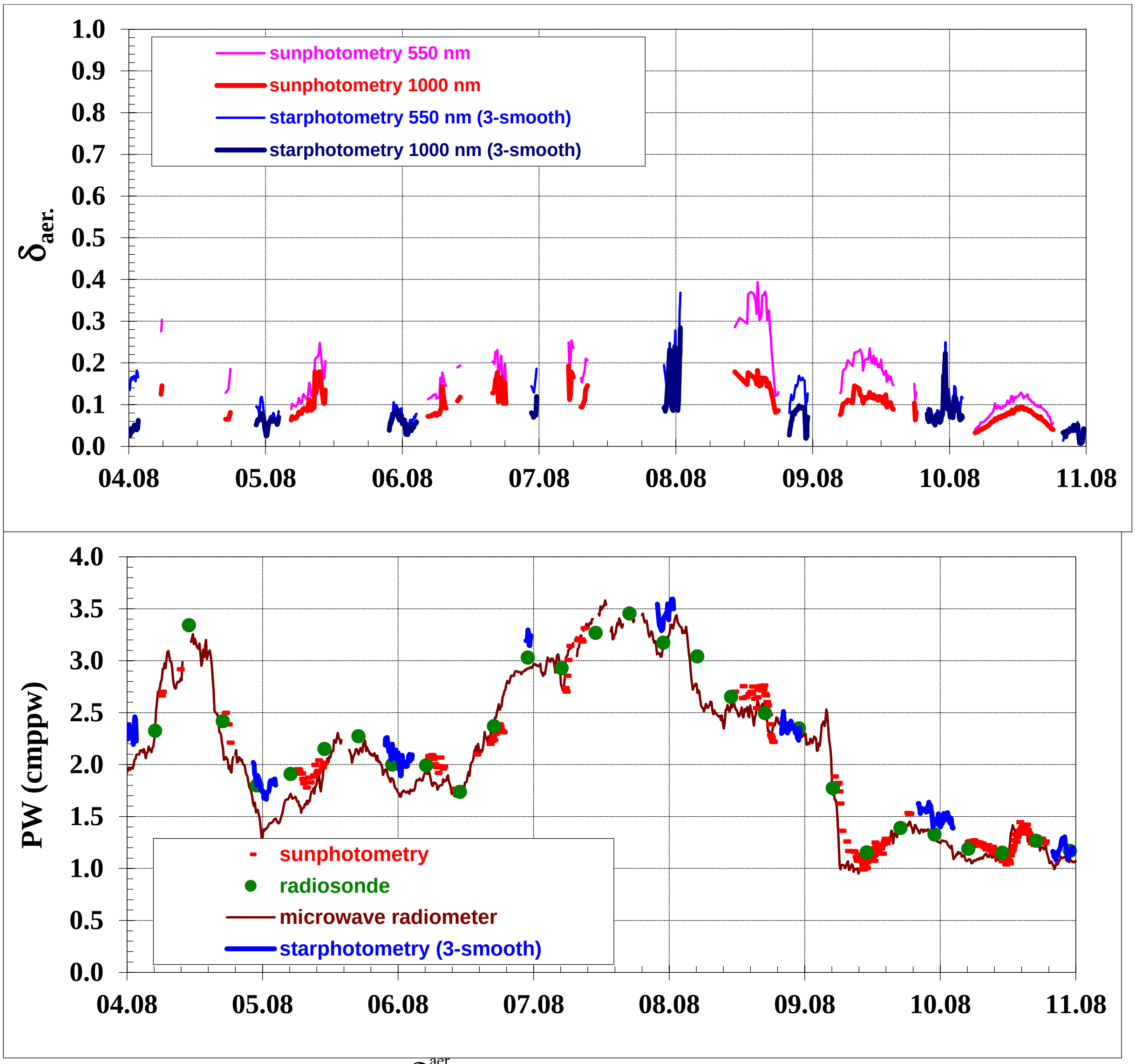

**Fig.5. Fragment of $\delta_\lambda^{aer.}$ - and PW-monitoring for 4-11 Aug 1998 (LACE 98).**

One can see the good compatibility of day and night data. For stable periods there is the tendency to decreasing of aerosol concentration in night time and to positive correlation between aerosol and water vapor concentration. The range of $\delta_\lambda^{aer.}$ - variations for stable periods was 0,2 – 0,4 and 0,0 – 0,2 for 0.55 **μm** and 1.0 **μm** accordingly. In general the amplitude of daily variations is smaller during the night than during the day due to convective processes.

The spectral behavior of aerosol was more complicate.

$$\delta_\lambda^{aer.} = A \cdot \lambda^{-B} \qquad \textbf{(12)}$$

As the rule the simple law (**12**) was applicable correctly only for spectral range ~ 0.44-0.8 **μm**. For **λ<0.4 μm** we observed frequently the lower aerosol concentration as well as for **λ>1.0 μm** the upper aerosol concentration was often observed. Also we often observed the relative minimum of aerosol concentration at **λ ~ 0.86 μm**. The example of different spectral aerosol distributions is shown on **Fig. 6** for 5 moments during day 20 July 1998. One can see that only for two moments (UTC 4:10 and 18:00) the law (**12**) are true (see coefficient of correlation $\mathbf{R^2}$). For other moments the law (**12**) is not applicable for whole spectral range. For all moment there is minimum of $\delta_\lambda^{aer.}$ near **λ ~ 0.86 μm**.

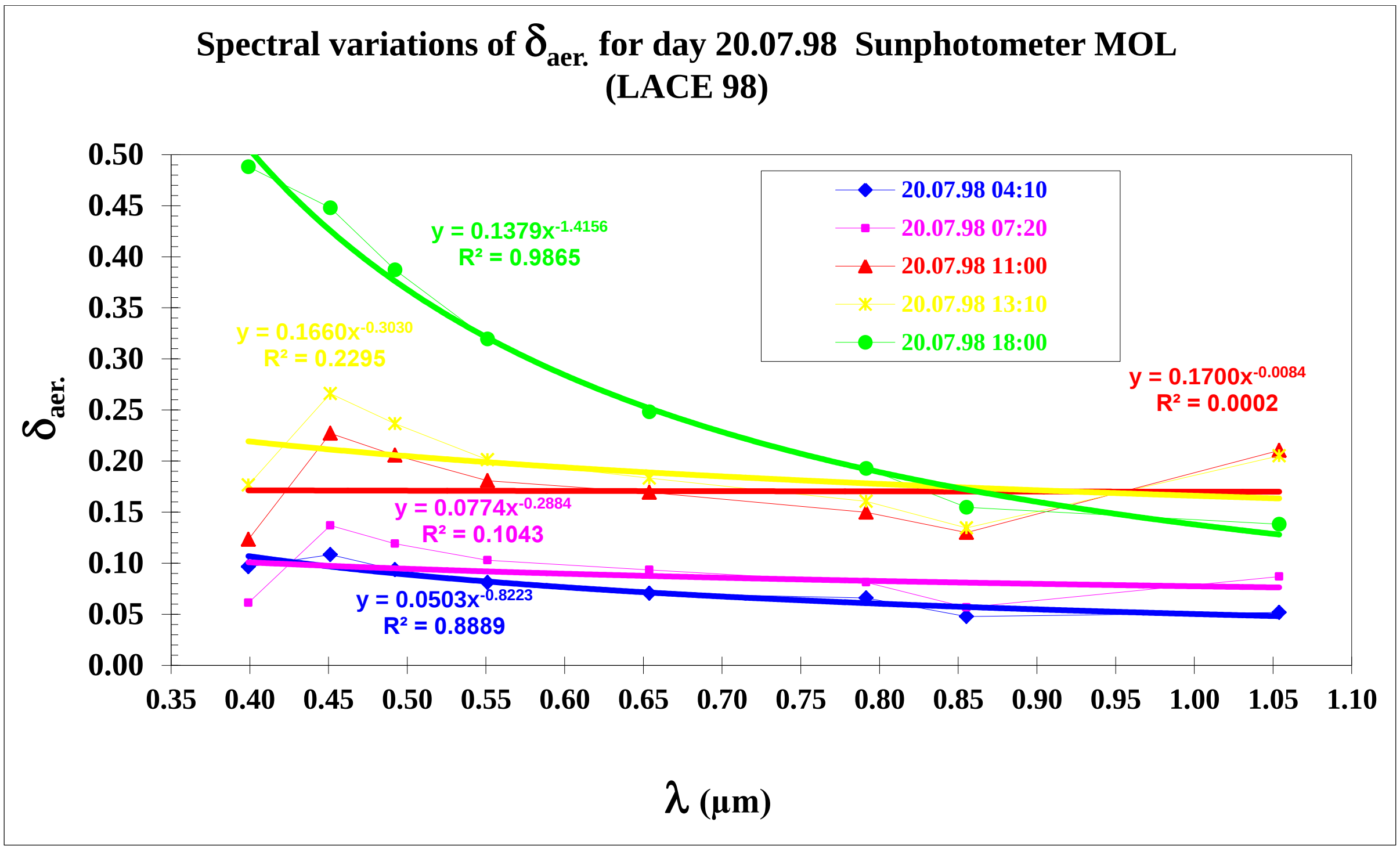

**Fig. 6. Complicate behavior of aerosol for day 20 Aug. 1998 (LACE 98).**

The most remarkable spectral effect we observed during night 4/5 August 1998 (see **Fig. 7**). During 30 min the $\delta_{\lambda}^{aer.}$ decreased in ~ 4-6 times, and whole spectral inversion was observed. It might be the downward adiabatic airmass transport process from stratosphere or upper troposphere layers to the ground.

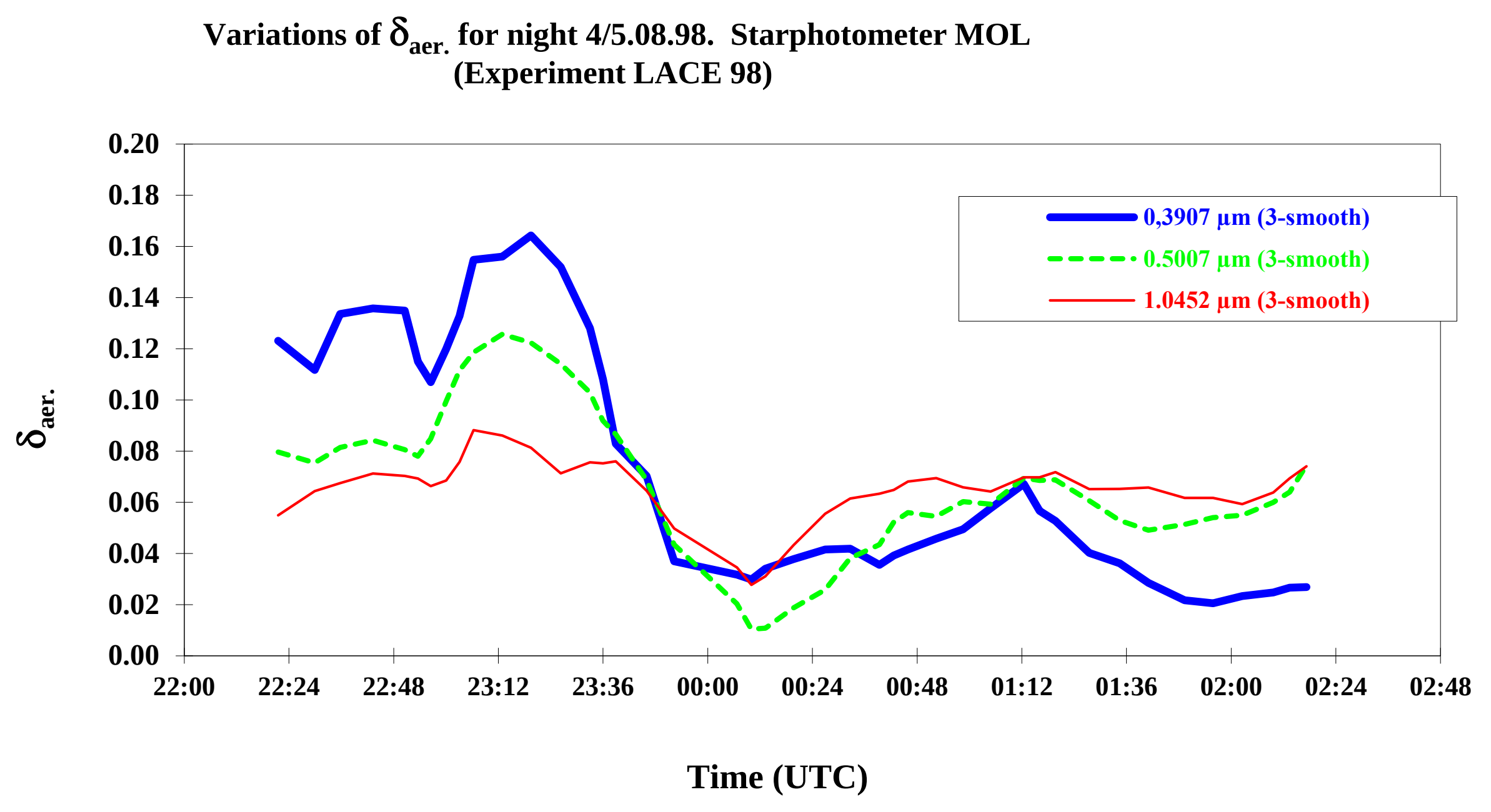

**Fig. 7. Remarkable aerosol behavior at night 4/5 Aug. 1998 (LACE 98).**

The our monitoring data for day and night 11/12 August 1998 was for the first time used successfully for validation of Raman Lidar measurements, **[10]**. The results of comparison are shown on **Fig. 8**.

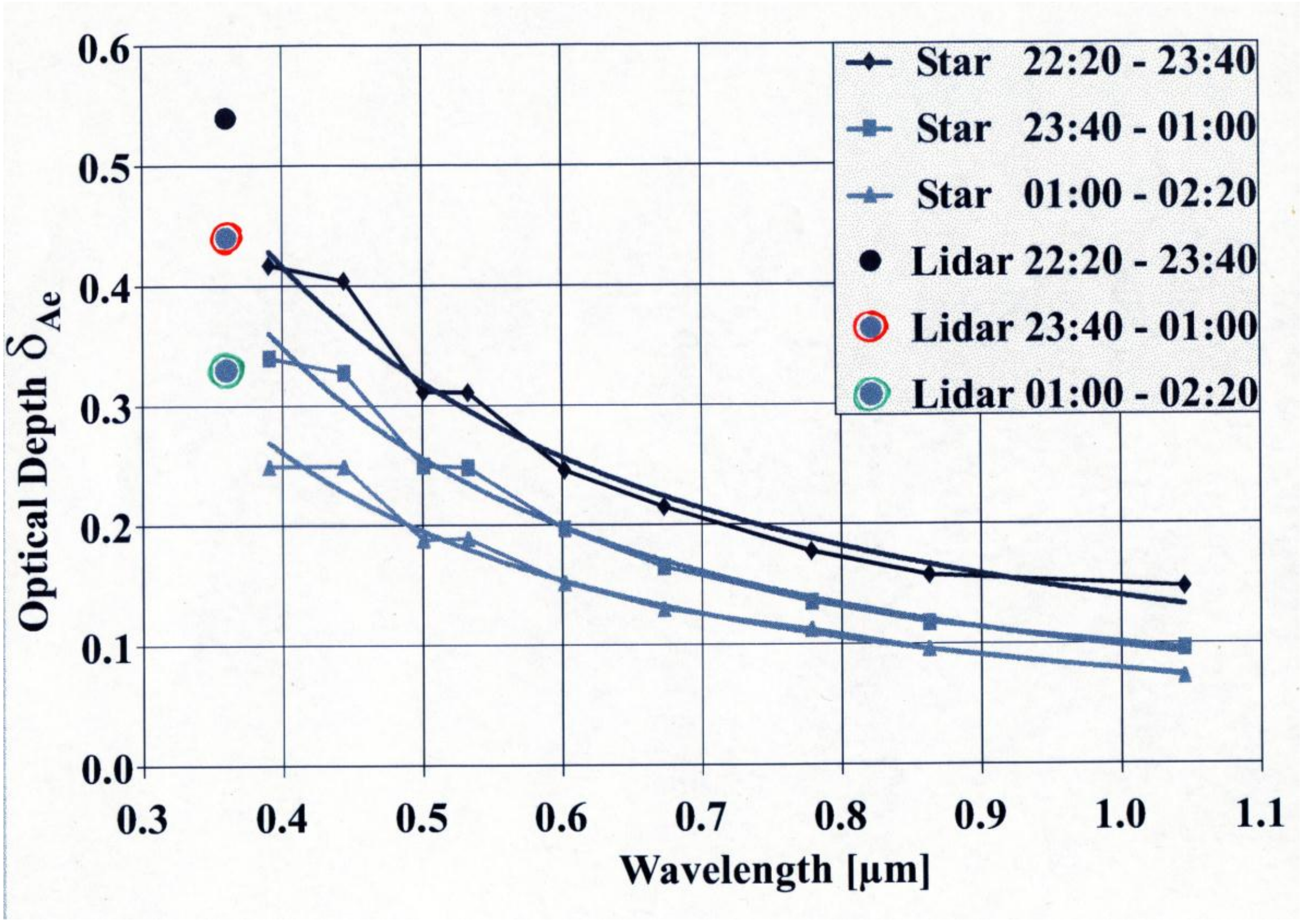

**Fig. 8. Comparison LIDAR's and photometric data (LACE 98).**

Due to good time-resolution of photometric measurements we could record very rapid (several minutes) PW-variations (see **Fig. 9**). The reality of its is verified by **MW**-radiometer's data.

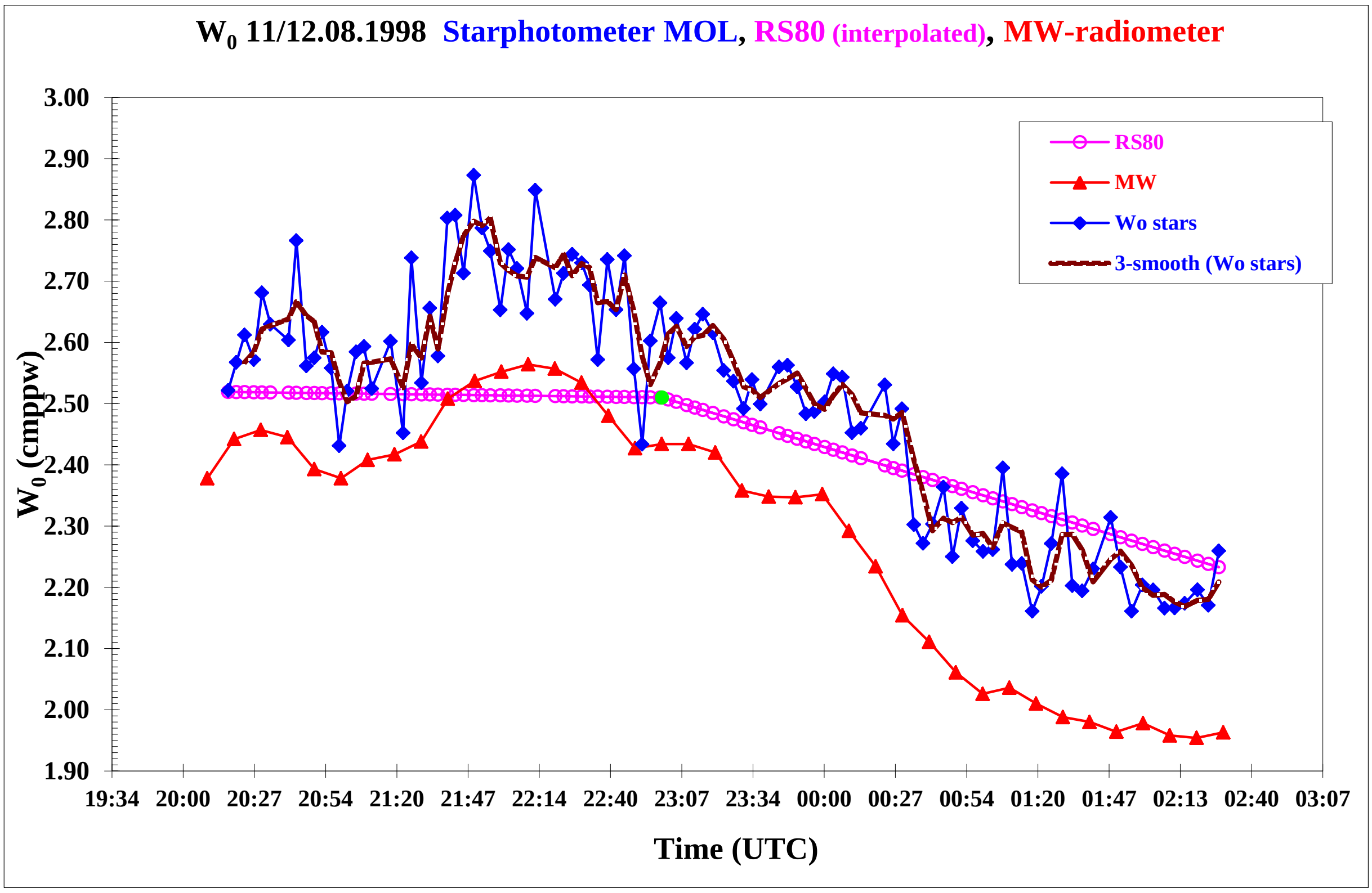

**Fig. 9. Rapid monitoring $W_0$ for night 11/12 August 1998 (LACE 98).**

## 3 Conclusions and Prospects

During LITFASS 98 and LACE 98 Experiments the optical method (sun- and starphotometry) was used for the first time as the equal in rights alongside with common geophysical methods. The large set of $\delta_\lambda^{aer.}$- and **PW**-monitoring data was received. Its analysis showed that this method allows to investigate in detail aerosol and water vapor behavior daily including the short-time-variations. Also for the first time the identical method of water vapor content determination was used both for solar and stellar observations.

All data showed good agreement between day and night observations. Also the water vapor content measurements showed the good agreements with results received by other independent methods (radiosondes and microwave radiometer).

For the first time sun- and starphotometric data were used successfully for calibration of aerosol **LIDAR**-measurements.

We consider the creation of full automatic version of starphotometer in MOL as the next aim. On this way we have the automatic starphotometer of Koldewey Station (Spitzbergen), **[11]**, as the good example.

We hope that sun- and starphotometric data will be used in future for calibration **LIDAR**- and satellite-measurements, and that optical method will become the common method for geophysical net practice.